\documentclass[12pt]{article}
\usepackage{amsmath,amssymb,amsfonts,amscd,bbm}
\begin{document} 


\thispagestyle{empty}
\renewcommand{\thefootnote}{\fnsymbol{footnote}}
\setcounter{footnote}{1}

\vspace*{-1.cm}
\begin{flushright}
OSU-HEP-03-15
\end{flushright}
\vspace*{1.8cm}

\centerline{\Large\bf Simple Model for (3+2)}
\vspace*{3mm}
\centerline{\Large\bf Neutrino Oscillations}

\vspace*{18mm}

\centerline{\large\bf 
K.S. Babu\footnote{E-mail: \texttt{babu@okstate.edu}} and 
Gerhart Seidl\footnote{E-mail: \texttt{gseidl@hep.phy.okstate.edu}}}
      
\vspace*{5mm}
\begin{center}
{\em Department of Physics, Oklahoma State University}\\
{\em Stillwater, OK 74078, USA}
\end{center}

\vspace*{20mm}

\centerline{\bf Abstract}
\vspace*{2mm}
We formulate a set of naturalness criteria for sterile neutrinos $(\nu')$ to be
light, needed for reconciling the LSND neutrino anomaly with the other neutrino data.
A light sterile neutrino becomes as natural as the light active neutrinos if
it carries quantum numbers of a chiral gauge symmetry broken at the TeV
scale. The simplest such theory is shown to be an $SU(2)$ gauge theory with the
$\nu'$ transforming as a spin 3/2 multiplet. We develop this model and show
that it leads naturally to the phenomenologically viable (3+2) neutrino
oscillation scheme. We also present next-to-minimal models for light sterile
neutrinos based on a chiral $U(1)$ gauge symmetry.


\renewcommand{\thefootnote}{\arabic{footnote}}
\setcounter{footnote}{0}

\newpage

\section{Introduction}
Recent solar \cite{SNO}, atmospheric \cite{superK}, and reactor \cite{Kam}
neutrino oscillation experiments have significantly improved our knowledge about neutrino masses and mixing angles. In particular, the solar and the atmospheric neutrino data are very well described in a three-neutrino oscillation scenario
where the mass squared splittings are
respectively $\Delta m^2_{\odot}\simeq 7.5\times 10^{-5}\:{\rm eV}^2$
and $\Delta m^2_{\rm atm}\simeq 2.0\times 10^{-3}\:{\rm eV}^2$ \cite{fogli}.
On the other hand, the $\overline{\nu}_\mu$--$\overline{\nu}_e$ oscillation
signal reported by the Liquid Scintillator Neutrino Detector (LSND) experiment
at Los Alamos \cite{LSND}, which will soon be
tested by the ongoing MiniBooNE \cite{MiniBooNE} experiment at Fermilab,
would require a third neutrino mass squared splitting
$\Delta m^2_{\rm LSND}\gtrsim 10^{-1}\:{\rm eV}^2$, which is impossible
to implement in a three-neutrino oscillation scheme. Instead, one
possibility to accomodate all the neutrino data is to add one or more
light sterile neutrinos with masses of the order $\sim 1\:{\rm eV}$, which would provide additional mass splittings. Although
four-neutrino mass models with a single sterile neutrino \cite{valle,bar}
are strongly disfavored by present data \cite{giun}, a  combined
analysis of the short-baseline experiments
Bugey \cite{Bugey}, CCFR \cite{CCFR84}, CDHS \cite{CDHS}, CHOOZ
\cite{CHOOZ}, KARMEN \cite{KARMEN}, and LSND shows,
that (3+2) neutrino mass schemes with two sterile neutrinos can yield a
satisfactory description of current neutrino oscillation data
including LSND \cite{sorel}. Generally, in (3+$n$) neutrino mass schemes,
where $n$ denotes the number of sterile neutrinos, it seems \cite{sorel} that
the LSND signal still remains compatible with the other data sets even
when $n>2$.

While the seesaw mechanism \cite{yana79} provides a simple understanding of the
smallness of active neutrino masses, it does not explain why a sterile
neutrino $\nu'$ would be light. In fact, if the effective low-energy theory is the Standard Model (SM), then there is no reason why $\nu'$ would not aquire a mass of the order of the Planck scale $M_{Pl}\sim 10^{19}\:{\rm GeV}$. Thus, in
any $(3+n)$ neutrino mass model it is important to explain the smallness of the sterile neutrino masses. In this paper, we wish to formulate a set of 
naturalness criteria for light sterile neutrinos
which would be as compelling as the seesaw mechanism for active neutrinos. We suggest and develop the simplest models which satisfy these criteria.

It is useful to recall the main ingredients that make the seesaw mechanism successful.
Here, the set of left-handed SM neutrinos\
$\nu_e,\nu_\mu$, and $\nu_\tau$ is extended by introducing three
right-handed neutrinos $N_1,N_2,$ and $N_3$, which are singlets under the SM
gauge group $G_{SM}=SU(3)_c\times SU(2)_L\times U(1)_Y$. In the basis $(\nu_e,\nu_\mu,\nu_\tau,N_1,N_2,N_3)$, the
resulting $6\times 6$ neutrino mass matrix then reads
\begin{equation}\label{eq:seesaw}
M_\nu=\left(
\begin{matrix}
0 & m_D\\
m_D^T & M_R
\end{matrix}
\right),
\end{equation}
where the entries $0$, $m_D,$ and $M_R$ are $3\times 3$ matrices which are characterized
by the gauge-structure and the Higgs-content of the theory. It is significant that,
in Eq.~(\ref{eq:seesaw}), the
entries in the upper-left $3\times 3$ sector are all
vanishing. This is because the SM is a chiral gauge-theory and does not permit a bare mass term for the left-handed neutrinos. In addition, there are no Higgs triplet fields, which could have directly coupled to $\nu_i$. Furthermore,
the matrix elements of $m_D$ are of the order of the electroweak scale
$\sim 10^2\:{\rm GeV}$ and
protected from becoming too large by electroweak gauge invariance. In contrast
to this, the mass matrix $M_R$ has unprotected entries of the order $M_{Pl}$
or of order the $B-L$ breaking scale $M_{B-L}\sim 10^{15}\:{\rm GeV}$.
As a result, we obtain an effective
$3\times 3$ neutrino mass matrix $M_\text{eff}=-m_DM_R^{-1}m_D^T$, which
leads to small neutrino masses of the order $\sim 10^{-2}\:{\rm eV}$.

By analogy with the seesaw mechanism for active neutrinos,
we propose the following criteria for a
light sterile neutrino $\nu'$ (with mass of order $1\:{\rm eV}$) to be natural:
\begin{enumerate}
\item $\nu'$ must transform as
a chiral representation of a ``sterile'' gauge symmetry
 $G'$ which is broken at the TeV scale.
\item There must exist no Higgs field which couples directly to $\nu'$.
\end{enumerate}
Note that we require $G'$ to be a gauge symmetry, rather than a global
symmetry, since only gauge symmetries will survive quantum gravity corrections.
In our constructions, we will supplement the above criteria by the requirement of a minimal Higgs sector: a single Higgs field breaks $G'$ and provides
simultaneously sterile neutrino masses, analogous to the SM Higgs doublet.

To illustrate the basic idea, let us consider the simplified case of
one generation with one active neutrino flavor $\nu$ and one sterile
neutrino $\nu'$ [(1+1) model]. Following our criteria, we extend the SM gauge symmetry to $G_{SM}\times G'$, with the $\nu'$ transforming chirally under $G'$.
All SM particles carry zero $G'$ charges.  Next, we introduce two
right-handed neutrinos $N$ and $N'$, which are singlets under the total gauge
group $G_{SM}\times G'$. In analogy with the electroweak symmetry breaking in the
SM, we assume that $G'$ is spontaneously broken around the TeV scale by
a suitable Higgs field $\Phi$ which has no direct Yukawa coupling of the type
$\nu'\nu'\Phi$. To keep the
situation simple, we furthermore take $\Phi$ to be a
singlet under $G_{SM}$. In the basis $(\nu,\nu',N,N')$, the total
$4\times 4$ neutrino mass matrix takes then the form
\begin{equation}\label{eq:sterileseesaw}
M_\nu=
\left(
\begin{matrix}
 0 & 0 &  m_{D} & m_{D}''\\
 0 & 0 &  \tilde{m}_{D}'' & m'_{D}\\
 m_{D} & \tilde{m}_{D}'' & M_{R} & M_{R}''\\
 m_{D}'' &  m'_{D} & M_{R}'' & M_{R}'
\end{matrix}
\right).
\end{equation}
 We hence observe, that the general principles which
lead to the usual seesaw mechanism, have also in this case
dictated the canonical structure of $M_\nu$
in Eq.~(\ref{eq:seesaw}).
Particularly, in Eq.~(\ref{eq:sterileseesaw}), the vanishing of the mass terms in the upper-left
$2\times 2$-block results from the chiral nature of the
$G_{SM}\times G'$ gauge theory and the absence of specific Higgs
representations. Moreover, $m_{D}$, $m_D'$, $m_D''$, and $\tilde{m}_{D}''$
are of the order $\sim 10^{2}\:{\rm GeV}$, since they are protected
by gauge invariance under $G_{SM}$ and $G'$ up to the TeV scale, where both
$G_{SM}$ and $G'$ are spontaneously broken. The entries $M_R$, $M_R'$, and
$M_R''$ on the other hand, are unprotected by $G_{SM}\times G'$ and thus of
the order $\sim M_{B-L}$. At low energies, this will therefore give
an effective $2\times 2$ neutrino mass matrix, which yields small masses
in the (sub-)eV-range for both the active and the sterile neutrinos.
The generalization of this sterile neutrino seesaw mechanism to a $(3+n)$
mass scheme is straightforward with $m_D$ becoming a $3\times n$ matrix and
$M_R$ becoming an $n\times n$ matrix in Eq.~(\ref{eq:seesaw}). Notice that in the special case when $G'$ is identified
with a copy of $G_{SM}$, we arrive at the well-known scenario for ``mirror'' neutrinos \cite{bere}. Alternative ways of realizing light sterile neutrinos have been suggested in Ref.~\cite{babu}.
 
In this paper, we construct the simplest neutrino mass model
consistent with our criteria for a light $\nu'$. As it turns out, the simplest model yields
the phenomenologically viable scenario of (3+2) neutrino oscillations
\cite{sorel}.
Here, we require invariance under the product group $G_{SM}\times G'$,
where $G'$ is a chiral anomaly-free continuous gauge symmetry. This implies, in particular, that no
extra discrete symmetry is imposed. The simplest example of this kind is found to be when $G'=SU(2)$, with the sterile neutrinos $\Psi$
in the spin 3/2 representation. A single
spin 3/2 Higgs field $\Phi$ can spontaneously break this symmetry at the
TeV scale without supplying large (TeV scale) masses to $\Psi$.
In this setup, we calculate the most general neutrino mass
matrix $M_\nu$ by explicitly minimizing the scalar potential for $\Phi$.
The minimum of the potential preserves a $Z_3$ subgroup of the sterile
isospin symmetry. The isospin $\pm\frac{3}{2}$ components of $\Psi$ are neutral under this $Z_3$, while the $\pm\frac{1}{2}$ components have charges $\pm 1$. Thus, only $\Psi_{\pm 3/2}$ will mix with the active neutrinos, yielding a (3+2) oscillation scheme. 
We also present the next simplest examples based on a chiral $U(1)$ gauge theory. Cancellation of chiral anomalies requires the existence of at least five
 -- more naturally six -- Weyl spinors, making these examples the second simplest.
 
\section{A simple chiral $SU(2)$ model}\label{sec:SU(2)}
The existence of a chiral gauge symmetry $G'$, broken at the TeV scale, plays a crucial r\^ole in our criteria for realizing naturally a light sterile neutrino. The vanishing of the axial vector anomalies and the mixed
gauge-gravitational anomalies \cite{alva} sets non-trivial constraints on such a theory. We are naturally led to the choice $G'=SU(2)$, where these anomalies automatically vanish for any representation. Furthermore, $SU(2)$ admits chiral representations, {\it i.e.}, fermionic representations for which mass terms
are forbidden by gauge invariance. Chiral $U(1)$ theories, while also interesting, are not the simplest as they require at least five spin 1/2 Weyl fermions for non-trivial anomaly cancellation. These next-to-minimal models are discussed in the next section.

Consider an $SU(2)$ gauge theory with one fermion field $\Psi$
in the spin $j$ representation. The spin $j$ representation of $SU(2)$
yields for $j=0,1,2,\dots$ (bosonic case) a unitary and for $j=1/2,3/2,5,2,
\dots$ (fermionic case) a projective unitary representation
(with essential cocycle) of $SO(3)$. Although the axial vector anomalies and
the mixed gauge-gravitational anomalies are zero for any $j$,
the spin 1/2 representation of $SU(2)$ is plagued with
a global Witten anomaly \cite{witt82}. The spin 1 representation will not suit our needs as it is vectorial.
The global $SU(2)$ anomaly vanishes, however, when
$\Psi$ transforms under the spin 3/2 representation, which has an even
quadratic index. In this case, $\Psi$ also cannot have an explicit mass term.
Therefore, $SU(2)$ with a single fermion $\Psi$ in the spin 3/2 representation is the simplest anomaly-free chiral gauge theory. $SU(2)$ with spin 3/2 matter fields has been studied in the context of dynamical supersymmetry breaking
in Ref.~\cite{intril94}. Non-Abelian chiral gauge theories are necessary ingredients for dynamical supersymmetry breaking and have been analyzed extensively
\cite{dine95}. 

The gauge symmetry of our model is $G_{SM}\times SU(2)$. We will assume here that the
$SU(2)$ symmetry is spontaneously broken by the vacuum
expectation value (VEV) of a single Higgs field $\Phi$ at the TeV scale. Like the fermion
$\Psi$, we put $\Phi$ into the spin 3/2 representation of the
$SU(2)$ symmetry. In component form, one can
write the $SU(2)$ spin 3/2 representations $\Psi$ and $\Phi$ as $\Psi=\left(\psi_1,\psi_2,\psi_3,\psi_4\right)^T$ and $\Phi=\left(\phi_1,\phi_2,\phi_3,\phi_4\right)^T$, where $\psi_1,\psi_2,\psi_3,$ and $\psi_4$ denote 2-component Weyl spinors and
$\phi_1,\phi_2,\phi_3,$ and $\phi_4$ are complex-valued\footnote{Recall that
the spin $j$ representation of $SU(2)$ is defined on the
space of polynomial functions on ${\mathbb{C}}^2$ that are homogeneous of
degree $2j$, which is a complex representation space.}
scalar fields. Here, we take all SM particles
to be singlets under the $SU(2)$ symmetry, while $\Psi$ and $\Phi$, on the other hand, are sterile with respect to $G_{SM}$. In addition, we assume seven
right-handed neutrinos $N_\alpha$ $(\alpha=1,\ldots ,7)$, which are total singlets under
$G_{SM}\times SU(2)$. For $n$ light $\nu'$ fields we will assume a total of
$n+3$ superheavy fields $N_\alpha$. As a result of the product group-structure and the fermionic charge assignment, this model is automatically free of all
anomalies.

The renormalizable Lagrangian relevant for neutrino masses is given by
\begin{equation}\label{eq:renormalizable}
 \mathcal{L}_Y=
a_{i\alpha}\ell_i HN_\alpha+b_{\alpha}\Psi\Phi^\ast N_\alpha
+c_\alpha\Psi\Phi N_\alpha+M_{\alpha\beta}
N_\alpha N_\beta+{\rm h.c.},
\end{equation}
where $H$ is the SM Higgs doublet, $\ell_i$ $(i=e,\mu,\tau)$ denotes the SM
lepton doublets, $a_{i\alpha},b_\alpha,$ and $c_\alpha$ are Yukawa couplings of order
unity and $M_{\alpha\beta}$ $(\alpha,\beta =1,\ldots ,7)$ are of order
$10^{14}-10^{16}\:{\rm GeV}$. Note that Eq.~(\ref{eq:renormalizable}) leads to
a mass matrix structure as given in
Eq.~(\ref{eq:sterileseesaw}). 
The effective dimension-five Lagrangian for neutrino masses is obtained after integrating out the $N_\alpha$
fields:
\begin{eqnarray}\label{eq:type1}
\mathcal{L}_{\rm eff}&=&\frac{H}{\Lambda}\ell_i\left[
Y_{1i}
\left(\psi_1\phi_1^\ast+\psi_2\phi^\ast_2+\psi_3\phi^\ast_3
+\psi_4\phi_4^\ast\right)
+Y_{2i}
\left(\psi_1\phi_4-\psi_2\phi_3+\psi_3\phi_2-\psi_4\phi_1\right)
\right]\nonumber\\
&+&\frac{Y_3}{\Lambda}
\left(\psi_1\phi_1^\ast+\psi_2\phi^\ast_2+\psi_3\phi^\ast_3
+\psi_4\phi_4^\ast\right)^2+\frac{Y_4}{\Lambda}
\left(\psi_1\phi_4-\psi_2\phi_3+\psi_3\phi_2-\psi_4\phi_1\right)^2\\
&+&\frac{Y_5}{\Lambda}
\left(\psi_1\phi_4-\psi_2\phi_3+\psi_3\phi_2-\psi_4\phi_1\right)
\left(\psi_1\phi_1^\ast+\psi_2\phi_2^\ast+\psi_3\phi_3^\ast+\psi_4\phi_4^\ast
\right)+
\frac{Y_{ij}}{\Lambda}H^2\ell_i\ell_j
+{\rm h.c.},\nonumber
\end{eqnarray}
where  $Y_{ij},Y_{1i},Y_{2i},Y_3$, $Y_4,$ and $Y_5$ $(i,j=e,\mu,\tau)$ are
dimensionless couplings related to $a_{i\alpha}$ and $b_{i\alpha}$
and $\Lambda\sim M_{ij}$. Here, the couplings $Y_{1i}$ and $Y_{2i}$, for example, arise respectively from the terms $\sim a_{i\alpha}b_\alpha$ and
$\sim a_{i\alpha}c_\alpha$ in Eq.~(\ref{eq:renormalizable}). The most general dimension-five neutrino mass operators which arise by integrating out arbitrary fermion representations
({\it i.e.}, by integrating out $SU(2)$ spin $j=1,2,3$ fermions in addition
to the $j=0$ states $N_\alpha$) are given in Appendix \ref{app:2ndtype}. These mass terms however, will not alter our general results here.

Following Appendix \ref{app:potential}, where the most general scalar potential for $\Phi$ has been minimized, we can assume a VEV of the form 
$\langle\Phi\rangle=(v_1,0,0,v_4)$, with $v_1$ and $v_4$ as given in
Eqs.~(\ref{eqs:extremum}) and $v_1,v_4\sim 10^2\:{\rm GeV}$. Since
$\langle\Phi\rangle$ breaks $SU(2)$
completely, the component-fields of $\Psi$ will finally appear as four sterile
neutrinos $(\nu_1',\nu_2',\nu_3',\nu_4')\equiv(\psi_1,\psi_4,\psi_2,\psi_3)$
in the low-energy theory (note in the definition the permutation of indices).

Integrating out the
right-handed neutrinos $N_\alpha$, the sterile neutrino seesaw
mechanism leads to five light
neutrinos with finite masses in the (sub-) eV-range and two massless
neutrinos. The massless states are $\nu_3'$ and $\nu_4'$ which decouple from
$\nu_e,\nu_\mu,\nu_\tau,\nu_1',$ and $\nu_2'$ (this is actually independent of
the total number of right handed neutrinos $N_\alpha$). The vacuum respects
an unbroken $Z_3$ symmetry, which is a subgroup of $I_3$, under which $\nu_3'$ and $\nu_4'$ have charges $\pm 1$ while the other fermionic fields are all neutral. This $Z_3$ symmetry forbids the mixing of $\nu_3'$ and $\nu_4'$ with the other neutrinos.
These states will aquire (sub-) eV masses once the effective Lagrangian
$\mathcal{L}'_{\rm eff}$ in Eq.~(\ref{eq:nonsinglet}) is taken into account.
The resulting
non-vanishing $5\times 5$ effective neutrino mass matrix can be written in the
basis $(\nu_e,\nu_\mu,\nu_\tau,\nu_1',\nu_2')$ as
\begin{equation}
 M_\text{eff}=\left(
\begin{matrix}
  \mathcal{M}_\nu & \mathcal{M}_\nu'\\
  {\mathcal{M}_\nu'}^T & \mathcal{M}_\nu''
\end{matrix}
\right),
\end{equation}
where $\mathcal{M}_\nu$ is an arbitrary $3\times 3$ matrix with entries of the
order $\sim10^{-2}\:{\rm eV}$, while $\mathcal{M}'_\nu$ is given by the $3\times 2$ matrix
\begin{equation}
\mathcal{M}'_\nu=
 \frac{\langle H\rangle}{\Lambda}
\left(
\begin{matrix}
 Y_{1e}v_1^\ast+Y_{2e}v_4 & Y_{1e}v_4^\ast-Y_{2e}v_1\\
 Y_{1\mu}v_1^\ast+Y_{2\mu}v_4 & Y_{1\mu}v_4^\ast-Y_{2\mu}v_1\\
 Y_{1\tau}v_1^\ast+Y_{2\tau}v_4 & Y_{1\tau}v_4^\ast-Y_{2\tau}v_1 
\end{matrix}
\right),
\end{equation}
and the $2\times 2$ matrix $\mathcal{M}_\nu''$ reads
\begin{equation}
 \mathcal{M}''_\nu=
\frac{Y_3}{\Lambda}
\left(
\begin{matrix}
{v_1^\ast}^2 & v_1^\ast v_4^\ast\\
v_1^\ast v_4^\ast & v_4^\ast
\end{matrix}
\right)-
\frac{Y_4}{\Lambda}
\left(
\begin{matrix}
-v_4^2 & v_1v_4\\
v_1v_4 & -v_1^2
\end{matrix}
\right)+
\frac{Y_5}{\Lambda}
\left(
\begin{matrix}
v_1^\ast v_4 & \frac{1}{2}(|v_4|^2-|v_1|^2)\\
\frac{1}{2}(|v_4|^2-|v_1|^2)&-v_1v_4^\ast
\end{matrix}
\right).
\end{equation}
It is therfore seen that
the effective interactions in Eq.~(\ref{eq:type1}) which generate the matrix
$\mathcal{M}_\nu'$ introduce a non-zero mixing of $\nu_1'$ and $\nu_2'$ with
the active neutrinos. Although the inclusion of the effective
operators $\mathcal{L}_{\rm eff}'$ in Eq.~(\ref{eq:nonsinglet}) lifts the
zero neutrino masses to small values of the order $\Lambda^{-1}v_1v_2\sim
 1\:{\rm eV}$, the
fields $\nu_3'$ and $\nu_4'$ will still remain decoupled from the rest of the
neutrinos, owing to the unbroken $Z_3$ symmetry. In total, the model therefore gives in any case a (3+2) neutrino
mass scheme for sterile neutrino oscillations.

\section{Simple chiral $U(1)$ models}\label{sec:U(1)}
In Sec.~\ref{sec:SU(2)}, we have analyzed a simple gauge extension of
$G_{SM}$ to $G_{SM}\times SU(2)$. It is instructive
to compare this model with a similar setup, where $SU(2)$ is replaced by a
sterile $U(1)$ gauge symmetry to give the total gauge group
$G_{SM}\times U(1)$. Let us therefore consider now
$N$ Weyl spinors $\Psi_{n_i}$ $(i=1,\ldots ,N)$, where $\Psi_{n_i}$ carries
the charge
$n_i$ under the $U(1)$ gauge group. In this model, the anomaly cancellation conditions read $\sum_{i=1}^Nn_i=0$ (mixed gauge-gravitational anomaly) and
$\sum_{i=1}^Nn_i^3=0$ (cubic gauge anomaly). It is easy to
see, that for $N\leq 4$ these conditions can only be fulfilled if the theory is
vector-like, {\it i.e.}, the $U(1)$ model must contain at least five
fermions to be chiral. Motivated by charge quantization, we shall require all
charges $n_i$ to be rational numbers, in which case they can be taken to be integers. Before discussing the case
of $N=5$ fermions, let us first consider simple chiral $U(1)$ models with
$N=6$. For this case, we find the following anomaly-free charge assignments:
\begin{subequations}\label{eq:N=6}
\begin{eqnarray}
{\rm Model\:(a)}\: :&2\times\{{\bf 5}\}+1\times\{-{\bf 3}\}
 +1\times\{-{\bf 2}\}+1\times\{{\bf 1}\}
 +1\times\{-{\bf 6}\},&\label{eq:2x5model}\\
{\rm Model\:(b)}\: :&2\times\{{\bf 4}\}+3\times\{-{\bf 1}\}
 +1\times\{-{\bf 5}\}.&
\end{eqnarray}
\end{subequations}
Here, Model (a), {\it e.g.}, has two Weyl fermions with $U(1)$ charge $5$ and one state each with charge $-3,-2,1,$ and $-6$. For Model (a), we minimally
extend the Higgs sector by adding a
single scalar singlet field $\Phi$ with $U(1)$ charge $-5$. From the charge assignment in Eq.~(\ref{eq:2x5model}) we then obtain the effective interaction 
Lagrangian for the neutrinos
\begin{equation}
\mathcal{L}_{\rm eff}=
\frac{1}{\Lambda}\ell_i H\Psi_5^\alpha\Phi+
\frac{1}{\Lambda}\ell_i\ell_j HH
+\frac{1}{\Lambda}\Psi_5^\alpha\Psi_5^\beta\Phi\Phi+\Psi_{-3}\Psi_{-2}\Phi^\ast+\Psi_1\Psi_{-6}\Phi^\ast+{\rm h.c.},
\end{equation} 
where $i=e,\mu,\tau$ and $\alpha,\beta=1,2$ and the Yukawa couplings have not
been explicitly displayed. Similar to the
$SU(2)$ model in Sec.~\ref{sec:SU(2)}, we suppose that $\Phi$ aquires its VEV at the TeV
scale. Hence, $\Psi_{1},\Psi_{-2},\Psi_{-3},$ and $\Psi_{-6}$ will
decouple below the TeV scale and we are left at low energies with a (3+2) model which is similar
to the $SU(2)$ model.

For Model (b), a minimal extension of the Higgs sector
by a scalar $\Phi$ with charge $-4$ leads to the effective neutrino mass
Lagrangian
\begin{equation}
 \mathcal{L}_{\rm eff}=
\frac{1}{\Lambda}\ell_i H\Psi_{4}^\alpha\Phi+
\frac{1}{\Lambda}\ell_i\ell_j HH
+\frac{1}{\Lambda}\Psi_{4}^\alpha\Psi_{4}^\beta\Phi\Phi+{\rm h.c.},
\end{equation}
where $\alpha,\beta=1,2$. This gives essentially a $(3+2)$ model with four additional extremely light neutrinos (the fields with charges $-1$ and $-5$) which decouple from the active
neutrinos. When $\Phi$, instead, carries the charge $+1$ we have the effective Lagrangian
\begin{equation}
  \mathcal{L}_{\rm eff}=
\frac{1}{\Lambda}\ell_i H\Psi_{-1}^\alpha\Phi+
\frac{1}{\Lambda}\ell_i\ell_j HH
+\frac{1}{\Lambda}\Psi_{-1}^\alpha\Psi_{-1}^\beta\Phi\Phi+
+\Psi_{4}^\gamma\Psi_{-5}\Phi+{\rm h.c.},
\end{equation}
where $\alpha,\beta=1,2,3$ and $\gamma=1,2$. These operators give rise to a
(3+3) scheme with one additional massless neutrino (a linear combination of $\Psi_4^\gamma$) and two heavy neutrinos ($\Psi_{-5}$ and one linear combination of $\Psi_4^\gamma$) which all decouple.

Let us now consider the case
of $N=5$ fermions. In Diophantine
analysis\footnote{This is a subject which is mainly concerned with the discussion of the rational or integer solutions of a polynomial equation $f(n_1,n_2,\ldots,n_N)=0$ with integer coefficients.} it has been shown that every integer
$n\neq\pm4\:(\text{mod}\:9)$ can be expressed as a sum of
the cubes of four integers \cite{mor69}. The integers $n=\pm 8\:({\rm mod}\:18)$, for example, can be written as
\begin{equation}\label{eq:expansion}
 (k-5)^3+(-k+14)^3+(3k-30)^3+(-3k+29)^3=18k+8\quad (k\in\mathbb{Z}).
\end{equation} 
Choosing in Eq.~(\ref{eq:expansion}) the value $k=28$, we arrive at the
integer solution $(n_1,n_2,n_3,n_4,n_5)\equiv (23,-14,54,-55,-8)$ of the
cubic anomaly cancellation condition. Note that none of the charges is
vector-like. Simultaneously, this solution also gives a zero mixed
gauge-gravitational anomaly.
As a result, the simplest anomaly-free chiral $U(1)$ theory with only
rational charges is given by
\begin{eqnarray}\label{eq:N=5}
{\rm Model\:(c)}\: :&
 1\times\{{\bf 23}\}+1\times\{-{\bf 14}\}+1\times\{{\bf 54}\}+
1\times\{-{\bf 55}\}+1\times\{-{\bf 8}\}.&
\end{eqnarray}
In comparison with the $N=6$ models (a) and (b) in Eqs.~(\ref{eq:N=6}),
however, the charges in Eq.~(\ref{eq:N=5}) involve rather large numbers, which makes this model less attractive.

\section{Discussion}
There are several experimental signatures of our models for naturally light
sterile neutrinos. Generally speaking, the most striking consequences will
be in the neutrino sector with very little effect  elsewhere.

First, a
confirmation of the LSND neutrino anomaly by MiniBooNE will clearly give
credence to this class of models. Second, since a (3+2) neutrino mass scheme
requires $U_{e5}\simeq 0.07$ \cite{sorel}, the model can be tested in the
future by $\overline{\nu}_e$ (or $\nu_e$) disappearence experiments. Moreover,
with a fifth neutrino mass eigenvalue $m_5$ in the range
$m_5\sim 4-6\:{\rm eV}$, the effective Majorana mass in neutrinoless double
$\beta$-decay $|\langle m\rangle|$ receives a contribution of the order
$\sim 0.02\:{\rm eV}$, which has a good chance to be tested in next
generation neurinoless double $\beta$-decay experiments like GENIUS, EXO,
MAJORANA, and MOON, which will have a sensitivity for
$|\langle m\rangle|\sim 0.01\:{\rm eV}$.

Due to the non-zero mixing of $H$ and $\Phi$, the SM
Higgs will have invisible decay modes such as $H\rightarrow \Phi\Phi$ and
$H\rightarrow W'W'$, if these decays are kinematically allowed. This can be tested at LHC or a future linear collider.

Clearly, the requirement $N_\nu<4$ on the total number of
neutrino species $N_\nu$ from ${}^4{\rm He}$ abundance in standard big bang
nucleosynthesis (BBN) \cite{wal} is violated,
since in all our schemes $\nu'$ will thermalize. However, there are
suggestions that a primordial lepton asymmetry will weaken this bound
\cite{foot}. Similarly, the neutrino mass limit
$\sum m_\nu < 0.7-1.0\:{\rm eV}$ $(@\: 95\%\:{\rm C.L.})$ from recent cosmological data \cite{el} may also be avoided for a suitable primordial $\nu_e$
chemical potential \cite{han}. Our viewpoint here is, that if the (3+2)
neutrino oscillation scheme is indeed confirmed by MiniBooNE, one will have to
revise the standard BBN paradigm.

Finally, it has been suggested that a
sterile neutrino in the 1--$20\:{\rm keV}$ range with very small mixing
 (${\rm sin}^2\theta\sim 10^{-11}$--$10^{-7}$ for $\nu'$-$\nu_e$ mixing) with the
active neutrinos can serve as a possible dark matter candidate and may be
responsible for the observed pulsar velocities exceeding
$\sim 500\:{\rm km}/{\rm sec}$ \cite{ful}. Our models are readily adaptable to
such a scenario.

\section*{Acknowledgments}
We would like to thank T. Schwetz for useful comments.  This
work is supported in part by DOE Grant \# DE-FG03-98ER-41076 and an award
from the Research Corporation.

\begin{appendix}
\section{Effective mass operators}\label{app:2ndtype}
Apart from the mass terms in Eq.~(\ref{eq:type1}), there exists in general a
second type of effective dimension-five neutrino mass operators, which arise 
by integrating out arbitrary fermionic $SU(2)$ representations.
The most general Lagrangain of these interactions reads
\begin{eqnarray}\label{eq:nonsinglet}
 \mathcal{L}_{\rm eff}'&=&
\frac{Y_6}{\Lambda}\left[ 2\left(\psi_1\psi_4-{\textstyle\frac{1}{3}}\psi_2\psi_3\right)
 \left(\phi_1\phi_4-{\textstyle\frac{1}{3}}\phi_2\phi_3\right)
 -{\textstyle\frac{4}{3}}\left({\textstyle\frac{1}{\sqrt{3}}}\psi_2^2-\psi_1\psi_3\right)
   \left({\textstyle\frac{1}{\sqrt{3}}}\phi_3^2-\phi_2\phi_4\right)\right.\nonumber\\
&-&\left.{\textstyle\frac{4}{3}}\left({\textstyle\frac{1}{\sqrt{3}}}\psi_3^2-\psi_2\psi_4\right)\left({\textstyle\frac{1}{\sqrt{3}}}\phi_2^2-\phi_1\phi_3\right)\right]+\frac{Y_7}{\Lambda}\left[
{\textstyle\frac{4}{3}}\left({\textstyle\frac{1}{\sqrt{3}}}\psi_2^2-\psi_1\psi_3\right)
   \left({\textstyle\frac{1}{\sqrt{3}}}{\phi_2^\ast}^2-\phi_1^\ast\phi_3^\ast\right)\right.\nonumber\\
&+&\left. 2\left(\psi_1\psi_4-{\textstyle\frac{1}{3}}\psi_2\psi_3\right)
\left(\phi_1^\ast\phi_4^\ast-{\textstyle\frac{1}{3}}\phi_2^\ast\phi_3^\ast\right)+{\textstyle\frac{4}{3}}\left({\textstyle\frac{1}{\sqrt{3}}}\psi_3^2
-\psi_2\psi_4\right)
\left({\textstyle\frac{1}{\sqrt{3}}}{\phi_3^\ast}^2-\phi_2^\ast\phi_4^\ast\right)\right]\nonumber\\
&+&\frac{Y_8}{\Lambda}\left[
{\textstyle\frac{2}{3}}\left({\textstyle\frac{1}{\sqrt{3}}}\psi_2^2-\psi_1\psi_3\right)
\left(\phi_1^\ast\phi_2+{\textstyle\frac{2}{\sqrt{3}}}\phi_2^\ast\phi_3+\phi_3^\ast\phi_4\right)+
\left(|\phi_1|^2+{\textstyle\frac{1}{3}}|\phi_2|^2-{\textstyle\frac{1}{3}}|\phi_3|^2 -|\phi_4|^2\right)\right. 
\nonumber\\
&\times&\left.
\left(\psi_1\psi_4-{\textstyle\frac{1}{3}}\psi_2\psi_3\right)
-{\textstyle\frac{2}{3}}\left({\textstyle\frac{1}{\sqrt{3}}}\psi_3^2-\psi_3\psi_4\right)
\left(\phi_1\phi_2^\ast+{\textstyle\frac{2}{\sqrt{3}}}\phi_2\phi_3^\ast+\phi_3\phi_4^\ast\right)\right]+{\rm h.c.},
\end{eqnarray}
where $Y_6,Y_7,$ and $Y_8$ denote Yukawa couplings of order unity.
The most general effective neutrino mass operators are thus given by the sum
$\mathcal{L}_{\rm eff}+\mathcal{L}'_{\rm eff}$. The gauge singlets in
Eq.~(\ref{eq:nonsinglet}) can be determined from a
Clebsh-Gordan table or by representing $\Psi$ as a totally symmetric tensor
$\psi_{ijk}$, where $i,j,k=1,2$ and the (normalized) components
are defined as $\psi_{111}=\psi_1$,
$\psi_{112}=\psi_{121}=\psi_{211}=\frac{1}{\sqrt{3}}\psi_2$,
$\psi_{122}=\psi_{212}=\psi_{221}=\frac{1}{\sqrt{3}}\psi_3$, and
$\psi_{222}=\psi_4$ (correspondingly for $\Phi$). In this notation, the
coupling $\sim Y_7$, {\it e.g.}, can be obtained from the term
$\psi_{abc}\psi_{ijk}\phi^{abi}\phi^{cjk}$ (summation of indices understood).

\section{Properties of the scalar potential}\label{app:potential}
 The most general renormalizable scalar potential of a $SU(2)$ spin 3/2 Higgs
representation $\Phi=(\phi_1,\phi_2,\phi_3,\phi_4)^T$
is given by
\begin{eqnarray}\label{eq:potential}
V &=&-\mu^2\left(|\phi_1|^2+|\phi_2|^2+|\phi_3|^2+|\phi_4|^2\right)
     +\lambda_1\left(|\phi_1|^2+|\phi_2|^2+|\phi_3|^2+|\phi_4|^2\right)^2
     \nonumber\\
  &+&\lambda_2\left(
    \left|{\textstyle\sqrt{\frac{2}{3}}}\phi_1\phi_3-
{\textstyle\frac{\sqrt{2}}{3}}\phi_2^2\right|^2
   +\left|\phi_1\phi_4-{\textstyle\frac{1}{3}}\phi_2\phi_3\right|^2
   +\left|{\textstyle\sqrt{\frac{2}{3}}}\phi_2\phi_4-{\textstyle
\frac{\sqrt{2}}{3}}\phi_3^2\right|^2   
\right)\nonumber\\
&+&\lambda_3\left[2\left({\textstyle\sqrt{\frac{2}{3}}}\phi_1\phi_3
-{\textstyle\frac{\sqrt{2}}{3}}\phi_2^2\right)\left({\textstyle\sqrt{\frac{2}{3}}}\phi_2\phi_4-
{\textstyle\frac{\sqrt{2}}{3}}\phi_3^2\right)-
\left(\phi_1\phi_4-{\textstyle\frac{1}{3}}\phi_2\phi_3\right)^2\right]\nonumber\\
&+&\lambda_4\left[
\phi_1^\ast\left(\phi_1^2\phi_4+{\textstyle\frac{2}{3\sqrt{3}}}\phi_2^3-\phi_1\phi_2\phi_3
\right)
+\phi_2^\ast\left(
\phi_1\phi_2\phi_4-{\textstyle\frac{2}{\sqrt{3}}}\phi_1\phi_3^2+{\textstyle\frac{1}{3}}\phi_2^2\phi_3
\right)
\right.\nonumber\\
&+&\left.
\phi_3^\ast\left(-\phi_1\phi_3\phi_4-{\textstyle\frac{1}{3}}\phi_2\phi_3^2+
{\textstyle\frac{2}{\sqrt{3}}}\phi_2^2\phi_4\right)
+\phi_4^\ast\left(
\phi_2\phi_3\phi_4-{\textstyle\frac{2}{3\sqrt{3}}}\phi_3^3-\phi_1\phi_4^2\right)
\right]+{\rm h.c.},
\end{eqnarray}
where the coefficients $\mu,\lambda_1,\lambda_2,$ and $\lambda_3$ are
real-valued\footnote{The phase of $\lambda_3$ can always be removed by an
appropriate phase-redefinition $\Phi\rightarrow e^{{\rm i}\varphi}\Phi$.} and
$\lambda_4=|\lambda_4|\cdot{\rm exp}({\rm i}\beta)$ with some arbitrary phase
$\beta$. Notice that the potential $V$ posesses the following
$U(1)$ symmetry which is part of the $SU(2)$ symmetry and
allows to set one phase of the fields $\phi_i$ always to zero:
\begin{eqnarray}
 U(1)&:&\phi_1\rightarrow e^{+{\rm i}\varphi}\phi_1,\quad
        \phi_2\rightarrow e^{+{\rm i}\varphi/3}\phi_2,\quad
        \phi_3\rightarrow e^{-{\rm i}\varphi/3}\phi_3,\quad
        \phi_4\rightarrow e^{-{\rm i}\varphi}\phi_4.
\end{eqnarray}
The potential $V$ has a local extremum of the form
$\langle\Phi \rangle=(v_1,0,0,v_4)$, where the complex entries $v_1$ and
$v_4$ have a relative phase $\alpha$, {\it i.e.}, it is
$v_1v_2=|v_1v_2|\cdot{\rm exp}({\rm i}\alpha)$. For simplicity, we may consider the limit 
$|\lambda_4|\ll 1$, in which case these quantities can be expressed to leading order as
\begin{subequations}\label{eqs:extremum}
\begin{eqnarray}
|v_1|^2&\simeq&\frac{\mu^2}{4\lambda_1+\lambda_2-2|\lambda_3|}
\left(
1\pm\frac{2|\lambda_4|}{\lambda_2-2|\lambda_3|}\:{\rm cos}(\beta)
\right),\\
\left|\frac{v_4}{v_1}\right| &\simeq&
1\pm\frac{2 |\lambda_4|}{\lambda_2-2|\lambda_3|}\:{\rm cos}(\beta),
\\
\alpha &\simeq&\frac{|\lambda_4|^2\:{\rm sin}(2\beta)}{\lambda_3(\lambda_2-2\lambda_3)}.
\end{eqnarray}
\end{subequations}
Notice in Eq.~(\ref{eq:potential}) that each interaction invlolves either zero, two, or four of the fields $\phi_2$ and/or $\phi_3$. In the minimum
$(v_1,0,0,v_4)$, the mixing of $\phi_2$ and $\phi_3$ with $\phi_1$ and
$\phi_4$ is hence zero. As a consequence, the mass matrix of $\phi_2$ and
$\phi_3$ has one pair of zero 
eigenvalues which correspond to two (would-be) Nambu-Goldstone bosons and two
degenerate non-zero mass-squared eigenvalues of the form
\begin{equation}
 m_{H^{\pm}}^2=\frac{2}{3}\left(|v_1|^2+|v_4|^2\right)
\left(\lambda_2+6\frac{|\lambda_4 v_1v_4|\:{\rm cos}(\alpha)}{|v_4|^2-|v_1|^2}\right).
\end{equation}
To calculate the remaining scalar masses, we consider the fluctuations
$\phi_1=v_1+\tilde{\phi_1}$, and
$\phi_4=v_4+\tilde{\phi_4}$ about the minimum $(v_1,0,0,v_2)$. The
corresponding mass eigenstates $G,A,H_1,$ and $H_2$ can be expressed as
\begin{subequations}
\begin{eqnarray}
 G&=&\frac{\sqrt{2}\:{\rm Im}(v_1^\ast\tilde{\phi_1}
-v_4^\ast\tilde{\phi_4})}{\sqrt{|v_1|^2
+|v_2|^2}},
\qquad
A\:=\:\frac{\sqrt{2}\:{\rm Im}(v_4\tilde{\phi_1}+v_1\tilde{\phi_4})}{\sqrt{|v_1|^2+|v_2|^2}},\\
H_1&=&\frac{\sqrt{2}\:{\rm Re}(v_1^\ast\tilde{\phi_1}-v_4^\ast\tilde{\phi_4})}
{\sqrt{|v_1|^2+|v_2|^2}},\qquad
H_2\:=\:
\frac{\sqrt{2}\:{\rm Re}(v_4^\ast\tilde{\phi_1}+v_1^\ast\tilde{\phi_4})}
{\sqrt{|v_1|^2+|v_2|^2}}.
\end{eqnarray}
\end{subequations}
The scalar $G$ is a massless (would-be) Nambu-Goldstone boson which has zero mixing with the other fields. In the limit $|\lambda_4|\ll 1$, the $3\times 3$
mixing matrix of the fields $A,H_1,$ and $H_2$ has the mass-squared eigenvalues
\begin{subequations}
\begin{eqnarray}
m^2_{1,2}&\simeq&
(2\lambda_1+\lambda_2)(|v_1|^2+|v_2|^2)\nonumber\\
&\pm&\sqrt{(2\lambda_1+\lambda_2)^2(|v_1|^2+|v_4|^2)^2
-8\lambda_1\lambda_2(|v_1|^2-|v_4|^2)},\\
m^2_3 &\simeq&+\frac{|\lambda_4|}{|v_1|v_4||}
(|v_1|^2+|v_2|^2)(|v_1|^2-|v_2|^2)\:{\rm cos}(\beta).
\end{eqnarray}
\end{subequations}
In total we see, that for a range of parameters the extremum described in
Eqs.~(\ref{eqs:extremum}) will be a local minimum. In this minimum, the
$SU(2)$ gauge symmetry is completely broken, thereby leaving three (would-be)
Nambu Golstone bosons, which must be eaten by the gauge bosons via the Higgs
mechanism.  The kinetic term of $\Phi$ is obtained from the covariant derivative
$$D_\mu\phi_{ijk}=\partial_\mu\phi_{ijk}-i\frac{g_2'}{2}\left[(W'_\mu)^\alpha_{i}
\phi_{\alpha jk}+(W'_\mu)^\alpha_{j}\phi_{i\alpha k}+(W'_\mu)^\alpha_{k}\phi_{ij\alpha}\right],$$ where $g_2'$ is the gauge
coupling and $(W'_\mu)^l_i$ $(i,l=1,2)$ are the $SU(2)$ gauge bosons.
In the minimum
$\langle\Phi\rangle=(|v_1|,0,0,|v_4|\cdot e^{{\rm i}\alpha})$, the gauge boson
masses are
\begin{equation}
m^2_{W_3'}=\frac{9}{2}{g_2'}^2(|v_1|^2+|v_4|^2)=3m_{W'_{\pm}}^2.
\end{equation}

\end{appendix}

\end{document}